%% file: 0_main.tex
\documentclass[runningheads]{llncs}
\usepackage{graphicx}
\usepackage{comment}
\usepackage{ifthen}
\usepackage{algorithm}
\usepackage{algpseudocode}
\begin{document}
\title{Fairness in ERC token markets: A Case Study of CryptoKitties}
%
%
\author{Kentaro Sako \and Shin'ichiro Matsuo \and Sachin Meier}
%
%
%
\maketitle              
\begin{abstract}
Fairness is an important trait of open, free markets.
Ethereum is a platform meant to enable digital, decentralized markets.
Though many researchers debate the market's fairness, there are few discussions around the fairness of automated markets, such as those hosted on Ethereum. 
In this paper, using pilot studies, we consider unfair factors caused by adding the program. 
Because CryptoKitties is one of the major blockchain-based games and has been in operation for an extended period of time, we focus on its market to examine fairness.
As a result, we concluded that a gene determination algorithm in this game has little randomness, and a significant advantage to gain profit is given to players who know its bias over those who do not.
We state incompleteness and impact of the algorithm and other factors.
Besides, we suppose countermeasures to reduce CryptoKitties' unfairness as a market.

\keywords{CryptoKitties  \and Smart contracts \and Financial market fairness.}
\end{abstract}
\input{1_introduction}
\input{2_fair_market_condition}

\input{3_vulnerability}

\input{4_countermeasures}

\input{5_future_work}

\input{6_conclusion}

\end{document}

%% file: 1_introduction.tex
\section{Introduction}
\subsection{Background}
After Bitcoin was proposed, many challenges are conducted to make economic activities performed autonomously without any trusted party. Bitcoin tries to realize such a space for a simple application like payment. On the other hand, with Solidity and another language, Ethereum tries to realize ``smart contract'' beyond the payment process. The movement is recently expanding decentralized finance. When we deal with the simple payment process, requirements on application-level security are a bit simple, preventing double-spending in the case of bitcoin.
The amount of payment is assumed to be correctly agreed among the payer and payee.
KYC/AML is the other regulatory requirement under debate.

On the other hand, in the case of a smart contract, such requirements become complicated.
Throughout our experience regarding Initial Coin Offering, there is a potential to scam due to asymmetric knowledge and some unfair situation for participants.
One of the significant expectations of permissionless block-chain mechanism and smart contract as its application is to provide transparency and fairness of an economic system. It may be true for payment applications like Bitcoin and cryptocurrency, but it is unknown if we can expect the same fruits for smart contracts.

Although there is a lot of research about the ordinary financial system's fairness, they do not discuss the fairness of markets run autonomously by programming code. When we try to discuss the fairness of smart contracts, we need to consider two aspects, at minimum,  addition to the concept of the fairness of the ordinary financial system; (1) effect by autonomous execution, and (2) trust of the programming code. Autonomous execution may make users difficult to manage their assets and strategy and understand their financial transactions are executed over a fair setting. The user should trust the programming code of the smart contract platform. Though the developers claim that the programming code is disclosed at GitHub for transparency, average users do not have enough capability to understand the code. As an example of supply chain risks, it is hard to prove the execution code is the same as the source code at GitHub repository.

At the time of writing of this paper, we do not have good criteria to evaluate if a specific smart contract platform/application is fair or not.  Though it is big research to discuss the fairness of smart contracts, it is worth conducting research on the source of the unfairness of smart contracts. This direction will be the basis of such evaluation criteria.



\subsection{Our Contribution}
This research discusses the potential unfairness of a market created by the smart contract.
For example, we analyze CryptoKitties, a blockchain-based game, and make the most use of smart contracts on Ethereum to evaluate if it is fair or not as a market.
According to \cite{CK_market}, its economic effect is more than forty million dollars. Thus, the existence of potential unfairness may lead to a question regarding legitimacy as a place to exchange cryptoasset.

We investigated the internal algorithm of CryptoKitties to determine the price of each kitty potentially. In particular, we focus on how an ERC-721 token is created in CryptoKitties. In this game, a kitty is produced as an ERC-721 token. We assume that each player's goal at this game is the player earns Ethereum by exchanging tokens and enjoying the kitty. The characteristics of a newly born kitty, an ERC-721 token, are determined by this game's gene algorithm.
If this algorithm is not fair, there is a risk that users will unfairly lose Ethereum. We also research trading tokens among the owner of kitties. It gives all users a chance to get kitties.

As a result of the research, we found that CryptoKitties does not satisfy some fair market conditions. The gene determination algorithm does not have qualified randomness, and it has a huge influence on the determination of characters of a newborn kitty. It is a source of asymmetry of knowledge. Only a person who knows the nature of the random function can predict a potential new kitty's characteristics.
Therefore, it is possible to guess which kitty produces the most valuable kitty. We found that only users, who know this bias and can buy kitties that give birth to valuable kitties, can earn more Ethereum. When a player tries to sell a kitty cheaper than the breeding fee, his revenue will be smaller than his cost.
Thus, the game has an unrealistic assumption on players' literacy; all people must have the ability to understand the algorithm.
This fact does not mean that all players have opportunities to gain profit. Moreover, currently, the auction format in CryptoKitties has information asymmetry by conspiring with a seller and a bidder. When the seller tells the bidder when his auction starts, it is difficult for other players to participate because the bidder makes a successful bid as soon as it starts. We indicate that CryptoKitties may be providing an unfair environment for many users.

In this paper, we argue conditions that a fair market should be kept in Section 2. In practice, we compare CryptoKitties and fair market conditions in Section 3. Section 4 mentions countermeasures that may make CryptoKitties fairer. We consider other vulnerabilities of CryptoKitties in Section 5. Finally, we conclude our research in Section 6.

\subsection{Related Work}
\paragraph{Alesja Serada et al.} He studied CryptoKitties as a subject to see how blockchain will shape future game design. He examined the relationship between token ownership and the value construction of CryptoKitties. In addition, he showed how the breeding and market aspects of kitty work concerning maintaining the game economy. As a result, the authors showed that the kitty's value is decreasing because there is no upper limit to the number of kitties that can be bred. The value of Gen $0$ kitties, which cannot be created by breeding, decreased as well. We also showed that the existence of a transaction fee GAS could hinder the intervention of new users. He concludes that these are the points that make the game economy unsustainable \cite{CK_new_ludic}.

\paragraph{Charlotte et al.} Based on that "trust without trust," blockchain has emerged as a disruptive technology that is considered an alternative to law. The authors doubt that whether participants can transact with each other without the need for legally sanctioned trust. The authors specifically highlight the need for users to verify that a Dapps (short for decentralized applications) really does fall under it. He focuses on some Dapps, including CryptoKitties. Since it is possible that CryptoKitties is not decentralized, it is marketing as a Dapp may be misleading to users. The reason why kitty is considered immutable and cannot be taken away from others is the blockchain's immutability. However, only the market uses its properties. Charlotte points out that it is a vulnerability, and some can cheat others to execute a dishonest contract \cite{CK_How}.

%% file: 2_fair_market_condition.tex
\section{Considering the Fairness of financial services based on smart contract}

\subsection{Preliminary}

\subsubsection{Blockchain}

Blockchain is a database commonly used as a ledger for cryptocurrencies. Satoshi Nakamoto proposed it as a bitcoin ledger in 2008 \cite{bitcoin}. Blockchain has some special characteristics; it enables decentralized systems, immutable data, transparency, and anonymity. 

A blockchain consists of many blocks. Each block contains transactions, a timestamp, a previous hash, and a nonce. A transaction has a sender address, a recipient address, and a value.

There is no administrator in the blockchain. Instead, every member of the network manages blockchain data. A Peer-to-Peer network connects the participants as nodes. Each node has blockchain data. If someone creates a new block, he sends all nodes connecting him to the block. These nodes will send other nodes when they receive the block. Soon, everyone will have that information. 

So, how do users make a new block? First, block creators called miners to determine a block which they want to connect their block. If they determine the transactions in that block are legitimate, they will make the previous hash of their block the hash value of that block. It is called a blockchain because the blocks are connected like a chain by hashing. If there are six or more blocks connected behind a block, it is considered correct. Next, miners select transactions which they thought right ones and each transaction fee is high. If a miner creates a new block, he is rewarded with new coins. In the Ethereum blockchain, the transaction fee is called "Gas." Then, miners calculate "nonce" so that a block hash value is less than the threshold. As a block has nonce, if they change its value, they will also alter the block hash value. This threshold is set so that miners can find a nonce in 10 minutes, making it difficult for multiple blocks to be created simultaneously. If a block is easily created, it is immediately assumed to be the correct one, and they will approve suspicious transactions. The threshold prevents this. Finding the nonce is called Proof-of-Work (PoW), and the process of making the block is called mining. In this way, blockchain is a decentralized system.

Other properties also meet. Because a block created once will be saved in every node's server, he has to attack all nodes if someone tries to tamper with it. As it is too difficult, all participants can not change blocks. Besides, they can see all blocks. So, blockchain has transparency, such as seeing which address a cryptocurrency originated from. By using this, we know an address' balance. Of course, other users can not steal its cryptocurrency, thanks to the UTXO system \cite{utxo}. However, bitcoin and Ethereum addresses have anonymity so that people can not figure out a real person who has the address.

\subsubsection{Smart Contract}
Nick Szabo proposed smart contracts in the 1990s \cite{Smart_Contract}. We define a contract in advance. Everybody can not change it once defined. When a person agrees with its definition, the contract is executed. His and the contractor's settlement will be run automatically. We need no third party to run this contract. Take a vending machine, for example; the pre-definition is the product's price and pictures displayed. By selecting a juice, it is correctly executed until settlement. If the input is the same, smart contracts must have the same output. For a given contract, if the input is the same, the result must be the same regardless of who performs it. If this is not the case, then different people can buy the same juice at different prices. This would make smart contracts unreliable and different from the concept.

Ethereum has implemented smart contracts for the first time. With writing pre-definition on blockchain, all users do not re-write it. After pre-definition, programming code Solidity runs contracts. Thanks to this system, we can exchange Ethereum safely.

\subsubsection{CryptoKitties}
\begin{figure}
\centering
\includegraphics[width=9cm]{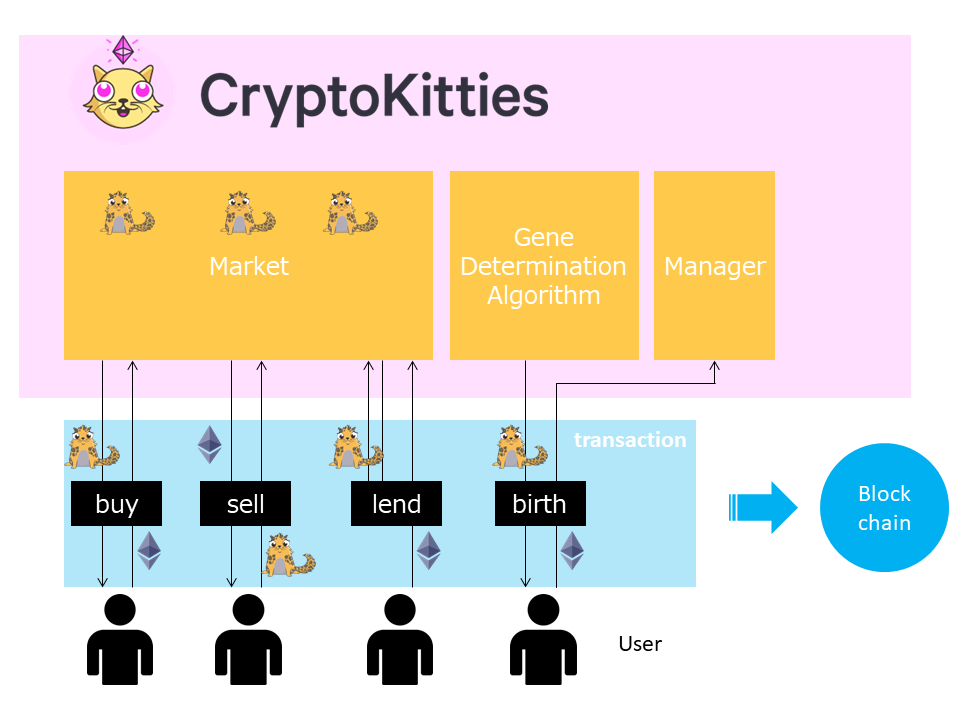}
\caption{CryptoKitties' Overview} \label{fig4}
\end{figure}

CryptoKitties is one of the most famous blockchain-based games \cite{CryptoKitties}. Axiom Zen created this game in 2017 \cite{CK}. We show CryptoKitties' overview in Fig \ref{fig4}. In this game, users exchange Ethereum and ERC-721 tokens. ERC-721 is a non-fungible token(NFT) transferred on the Ethereum blockchain. Unlike cryptocurrency tokens, NFTs are unique tokens, with specific parameters. Each ERC-721 token differs in its value. In CryptoKitties, an ERC-721 token is a kitty. Again, players exchange kitties and Ethereum. These kitties have an ID, gene, and generation. A kitty's ID is assigned in the order of birth, and the algorithm written in Solidity determines the gene that determines the appearance of the kitty. The generation of a child kitty is one greater than the generation of the parent kitties. By birthing and trading kitties, users aim to earn Ethereum. This game's source code is written in Solidity. So, all trades and breeding in this game are executed by smart contracts. Not only transactions but kitties' data are on the blockchain.

There are two ways to obtain a kitty. One is winning an auction. The auction of CryptoKitties is the dutch system that the exhibited kitty's price goes down as time passes. There are two types of auctions; standard and rental. When a user wins a standard auction, he can get a kitty. In case of rental, a winner has to return the kitty after breeding, another way to get a kitty.

A user can get a new kitty created by the gene determine algorithm. To make a kitty, he needs to choose two parent kitties that are inputs of the algorithm. Parent kitties can be chosen from his own kitties, or one of them can be a kitty he won at a rental auction. After selecting parent kitties, he can get a baby kitty. Thus, this process is called breeding. A player has to pay 0.008 Eth when he lets two kitties breed. Since a kitty has no gender, it can be either a matron or a sire. The breeding defines a baby kitty's gene and generation. A kitty's generation settles its breeding period \cite{CK_code} \cite{CK_cooldown}.
It has 14 kinds; the longest is two weeks, and the shortest is one minute. When a kitty is created, the cooldown period is determined by its generation, and from that point on, each time it is bred, the period increases by one kind. After breeding, until this period of time has passed, a matron kitty cannot reproduce. Through this kind of trading and breeding, players can get expensive kitties and sell them for a profit.

\subsection{Fairness in CryptoKitties Market} 

In terms of economics, a fair market should keep the following criteria \cite{criteria}. According to \cite{criteria}, every player has opportunities to profit and take risks equally. It must also be impossible to cheat to earn money. The information asymmetry, where some people know information about making a profit, must remain relatively small. The trading environment must also be equal for all participants. Finally, fairness includes adopting some measures to protect the weak.

We apply these requirements to CryptoKitties. We consider CryptoKitties' opportunities and information to gain profit, cheating, trading, and the inferior. To begin with, getting and selling a high-value kitty is the way to make Ethereum. So, an equal chance in this game means that all players can get high-value kitties. If you get a valuable kitty, you can sell this kitty and earn a lot of Ethereum.
Then, we assume CryptoKitties' cheating. It is earning Ethereum unfairly. Since players need to get kitties to earn, we focus on the way to obtain a kitty. Now, there are two methods to get a kitty: winning an auction or breeding. So, cheating could be winning the kitty without following the rules at the auction or obtaining it by tampering with the breeding algorithm. CryptoKitties must not allow either action.

Information about kitties is essential for players to maintain a relatively symmetric market. For example, what kind of kitty is being sold and at what price? How is a new kitty created? CryptoKitties should make all such information available. 
Next, we check the trading environment. Again, players can get a kitty by auction or breeding. It is the only auction that a player trades his kitty with other players. Therefore, there should be a rule of auction so that no one has a disadvantage.
Finally, we define the weak. According to \cite{criteria}, it takes players who do not have enough information, ability to negotiate and judge as examples. We have already defined information as about a kitty. Then, a player's bargaining power has to do with how valuable kitties he can get to gain profit. If he has a lot of Ethereum, he will get many kitties and valuable ones. Thus, this ability is related to financial resources. 
Besides, as mentioned in Section 2.1, the gene determination algorithm expresses how it creates a kitty. If a player understands this game's algorithm, he knows and can judge which kitty he should get and how kitties he should select as parents to gain profit. With understanding this game's algorithm and making the right decisions, he can make money, so judgment is affected by how well he understands this game. He will try to understand the algorithm written in Solidity to obtain an expensive kitty. Hence, the ability to judge is related to the ability to read Solidity. Therefore, we define the weak as players who have little Ethereum and can not read Solidity.

We found that CryptoKitties does not meet some of condition that we show below. Condition No.1 and No.2 are rules for protecting the socially vulnerable. Other states provide opportunities to gain profit for all users. We point out that this game does not meet all conditions except for No.4 in Section 3. In Section 5, we mention that CryptoKitties may not be satisfied with Condition No.1, 3, and 4 as a future work.

\begin{enumerate}
  \item There is a rule that avoids disadvantages for those who do not have large amounts of Ethereum.
  \item There is a rule that avoids disadvantages for those who can not understand Solidity.
  \item Information about the prices that are available in the market for each digital kitty is communicated instantaneously and costlessly to all users so that they know what trading opportunities exist.
  \item The users are sufficiently small compared to the size of the market that their supply and demand behavior does not have enough influence on the market to be recognized as such.
  \item The possibility of realizing profit opportunities is equally open to all users.
\end{enumerate}

%% file: 3_vulnerability.tex
\section{Analysis on CryptoKitties and its impact to fairness}
\subsection{Analysis on Gene Determination Algorithm}

Gene, one of a kitty's parameters, is a 240-bits number and depends on Gene Determine Algorithm, as shown in Algorithm \ref{algo1}. Again, gene defines a kitty's appearance. In detail, each of the five bits determines an element of appearance. For example, the ninth five bits correspond to the kitty's eyes' color. This algorithm is a smart contract and determined by the mixGenes function defined in GeneScience.sol \cite{CK_code}. 
In this algorithm, a gene array is used. Its length is 48, and each cell corresponds to an element of the kitty's appearance. The first cell is the last 5 bits of gene value. The second cell is the second last 5 bits of gene value. In the same way, determine all of the cells of the gene array. We elucidate whether the gene determination algorithm that builds the value of tokens to be traded in the market makes the market unfair.

\begin{algorithm}
\begin{algorithmic}[1]

  \State{$matron :=$  matron gene array}
  \State{$sire :=$ sire gene array}
  \State{$child :=$ child gene array}
  \State{$hash :=$ SHA-256(target block), $hash[i]$ means i-th bit of $hash$}
  \State{$k = 0$, k uses for $hash$}
  
  \For{$i = 0 \, \ldots \, 11$}
    \For{$j = 2 \, \ldots \, 0$}
      \If{$hash[k:k+2] == 0$}
        \State{$swap(matron[i*4+j],matron[i*4+j+1])$}
      \EndIf
      \State{$k+=2$}
    \EndFor
  \EndFor
  \For{$i = 0 \, \ldots \, 11$}
    \For{$j = 2 \, \ldots \, 0$}
      \If{$hash[k:k+2] == 0$}
        \State{$swap(sire[i*4+j],sire[i*4+j+1])$}
      \EndIf
      \State{$k+=2$}
    \EndFor
  \EndFor

  \For{$i = 0 \, \ldots \, 47$}
    \State{$mutated = false$}
    \If{$i \% 4==0$}
      \If{$abs(matron[i]-sire[i])==1$ and $min(matron[i],sire[i]) \% 2==0$}
        \If{$hash[k:k+3]<=1$}
          \State{$child[i] =$ $smallT/2$ $+ 16$}
          \State{$k+=3$}
          \State{$mutated$ $=$ $true$}
        \EndIf
      \EndIf
    \EndIf
    \If{$!mutated$}
      \If{$hash[k]==1$}
        \State{$child[i]=matron[i]$}
      \Else
        \State{$child[i]=sire[i]$}
      \EndIf
      \State{$k+=1$}
    \EndIf
  \EndFor \\
  \Return{child}

\end{algorithmic}
\caption{Gene determination algorithm} \label{algo1}
\end{algorithm}

The first step in this algorithm between lines 6 and 21 is a swap for parents' genes. Before the swap, the gene value is divided into twelve groups. Each group has four cells; the first four cells are group 0, the second four cells are group 1, and so on. Within each group, the swap operation is to change the order of the gene array. Next, the SHA-256 hash value of a block on the Ethereum blockchain, called the "target block," is involved in the algorithm. Six bits of the hash value of the target block are used for the swap in one group. Group 0 uses the last six bits, and group 1 uses the second last six bits and same as below. Let us say the four cells in the group are $a_0$, $a_1$, $a_2$, and, $a_3$ starting with the one with the smallest index. If the last two bits among six bits are both zero, $a_2$ and $a_3$ are swapped. Then, if the next two bits are both zero, $a_1$ and $a_2$ are swapped. Finally, if the remaining two bits are both zero, $a_0$ and $a_1$ are swapped. 

The next operation between lines 22 and 41 is to fill the cells of the gene array of the kitty to be born one by one (for $i=0$ to $47$). There are two methods for genetic determination in a new kitty; inheritance and mutation. The first action executed in this operation is checking to see if cell $i$ of the parents meets the requirements for mutation. If two cells are satisfied with 1, 2, and either 3-a or 3-b in below, the formula (1) determines the child gene's cell $i$. In that formula, smallT means the smaller of the two parents' cell $i$.

\begin{description}
  \item[1] $i$ is multiple of 4
  \item[2] the absolute value of matron cell $i$ and sire cell $i$ is 1, and smaller one is even
  \item[3-a] the smaller cell value is less than 22, and the lower three bits of the unused bits in the hash value of the target block are $001$ or $000$
  \item[3-b] the smaller cell value is not less than 22, and the three bits of the unused bits in the hash value of the target block are $000$.
\end{description}

\begin{equation}
  cell = smallT/2 + 16
\end{equation}
When any of the conditions are not met, a baby inherits either parent's cell. If the lowest bit of the unused bits in the hash value of the target block is $1$, cell $i$ of a child gene will inherit from the matron's gene. If not, it inherits from the sire.

In short, a baby gene is dependent on its parents' gene and target block hash. All kitties' genes are on the blockchain. By using Etherscan, a block explorer, we can check any kitty's genes. Then, what is the target block? We can predict a breeding result if we know the hash value. If it means randomly choosing one of all the blocks on the blockchain, we can not infer the outcome. However, the target block is somewhat limited and predictable. The target block is a block that will be issued when a matron kitty becomes fertile again. Specifically, there is a variable that stores the frequency of block creation. The product of that value and the matron kitty's breeding period corresponds to the blocks issued when she can breed again. We can find out the breeding period of the kitty from CryptoKitties' official page. Therefore, we know when the breeding period ends, and the block issued at that time or thereabouts becomes the target block. By calculating the hash value of the target block, we can predict a baby gene.

The fact that the results are predictable means that CryptoKitties has not satisfied conditions 1 and 5 of fairness, as defined in Section 2. First, the blockchain's transparency allows us to see what kitties are traded at a high price. Since we can expect breeding results, it is also possible to predict parent kitties to produce ones that match these trends. If we can make a successful bid to them, we will get a good kitty and earn Ethereum.

\begin{figure}
\centering
\includegraphics[width=3cm]{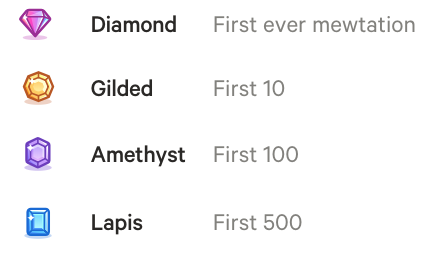}
\caption{Family Jewels} \label{fig2}
\end{figure}

However, in this kind of competition, well-financed users will have an immense advantage. We prove this by setting up a simple environment for CryptoKitties. First of all, we use "Cattribute," which means the attribute of kitty. In this market, there are 324 kinds of Cattribute. Let us take "driver" one of the Cattribute as an example. As of Jan. 13, there were 24 "driver" kitties in the market, and when we checked the gene sequences of all of them, we found that No.0 is 15 and No.36 is 23 in common. Therefore, we expect that the kitty that satisfies these two points has a "driver." In case of other Cattribute, "dominator," we found that No.0 is 28 and No.28 is 23 in common. Also, for a Cattribute, the first kitty to belong to that it gets Diamond. The first 10 kitties to belong to it will get Gilded, and the first 100 kitties will get Amethyst. From then on, these jewels are called "Family Jewels," as shown in Fig \ref{fig2} \cite{CK_family_jewels}. According to \cite{CK_family_jewels_amount}, the minimum price for each Jewel 5, 0.5, 0.07, 0.009 Eth, in order of rarity. Since these prices are positively correlated with rarity, the environment we set up should be like that. So, our environment will adopt these prices. In other words, all kitties with Diamonds are assumed to be 5 Eth. For the other kitties, we take those kitties that can produce $X$ Eth will be sold for $X/2$ Eth. For example, a kitty that can reliably give birth to a Diamond kitty would be priced at 2.5 Eth. Also, the minimum price of kitty to be sold in the market shall be 0.004 Eth. This amount is the minimum in the market as of Jan. 13.

The key is a kitty who has Diamond, which represents the pioneer of that Cattribute. Assuming that the value remains the same as more kitties of the same Cattribute, if a player gets a Diamond kitty, his earnings will be greater. Specifically, he gives birth to 499 kitties so that all of them inherit Diamond kitty's Cattribute and gets family jewels. He will get 9 Gilded, 90 Amethyst, and 400 Lapis kitties. Selling all kitties, his earnings will be 14.4 Eth. ($0.5*9 + 0.07*90 + 0.009*400$) Thus, the player who has a Diamond kitty will gain a lot of profit. Then, in order for this to happen, you need to win the auction or mutate and give birth to the Diamond kitty.

This algorithm will make it easy to know which kitty to get to make money. Such kitty is in high demand, and users with financial resources are more likely to win its competition. Alice, who has little Ethereum, can not buy Diamond kitty because it is expensive. So, to get these kitties, she has to birth them from other kitties. She can expect which kitties will give birth to a Diamond kitty. However, it is difficult for her to get these parent kitties because they are also in high demand. Since we can predict what kind of kitty will be born, some players may repeatedly give birth to make a parent kitty. But, Alice can not do a lot of births because a fee of 0.008 Eth is charged for breeding, and again she has little Ethereum. For Alice to make a profit, she needs to do things differently. It is a way of making a lot of small gains. However, since the birth fee is 0.008 Eth, selling a normal kitty will negatively affect revenue. So, she ought to sell kitties that have family jewels. As with the Diamond kitty, they are in high demand, and Alice will be hard to get them. Therefore, since the profit opportunity for players who do not have a lot of Ethereum is small, CryptoKitties violates Condition No.1 and No.5.

\subsection{The readability of Solidity source code}

Based on the previous section, rich players can make money, but is it really possible for everyone? Next is focusing on the actions they will take. The most important thing is understanding the gene determination algorithm. When they try to give birth to a Diamond kitty, they must find a Cattribute that has not been found yet. Besides, they have to choose parent kitties so that a Diamond kitty is created. In other cases, when trying to produce other kitties with family jewels from a Diamond kitty, they also have to select parent kitties so that the child inherits a parent's Cattribute. So, to make money, they need to understand the algorithm's behavior.

However, not everyone will be able to understand this algorithm. The main reason is that it is only written in Solidity. People who lack this Solidity knowledge will find it almost impossible to understand the algorithm. There is a considerable gap between those who can understand Solidity and those who can not, which significantly affects profit opportunities. Therefore, the results are contrary to Condition No.2 and No.5.

\subsection{Auction in CryptoKitties}

In this game, trading kitty is always done through auctions. So, users can only get kitties from the auction market. In this auction, a seller decides the starting and ending price of a kitty and auction. He can then start the auction with his intentions. According to condition No.3, a fair market should provide information about a trade for all players. It needs an environment where they know what kind of transactions exist. In this sense, information on what type of kitty is on sale and what price must be shared with all users.

However, when an exhibitor, Alice, and a user, Bob, are colluding, other players, Charlie, have little chance to get Alice's kitty. Alice tells when her auction tries to start Bob. As soon as her auction starts, Bob bids for her kitty, and Charlie can not see its trading. For example, Alice gives Bob a Diamond kitty in this way. Other players can not get her kitty, and Bob can get it and its child kitties. They have the same Cattribute as their parent's one and with family jewels. As a result, only Bob can gain profits. It contradicts condition No.3 and also enhances to expand information asymmetry.

%% file: 4_countermeasures.tex
\section{Enhancement of fairness}
\subsection{Gene Determination Algorithm}
As mentioned in Section 3.1, we state that it is difficult for not rich players to gain profit in CryptoKitties. To earn Ethereum, players need to sell a valuable kitty. Users with financial resources have the advantage of getting that kitty at auction. The method of giving birth and getting a kitty is also easier for wealthy users than ordinary ones. Because the outcome of the gene determination algorithm is predictable, so the demand for parent cats is high, and only rich people can afford them.

If the gene determination algorithm's output is unpredictable, everyone may have a chance to gain profit. With that algorithm, no one would know which parent kitties would produce a valuable kitty. All players do not know which kitties as parent kitties will make a profitable kitty. As a result, the value of the parent kitties that produced a valuable kitty is unknown. Therefore, there is a possibility that even a cheap kitty can give birth. If kitties are reasonable, many users can bid on them, so we think the game will be fairer than it is now.

However, there is a problem with introducing an unpredictable system in CryptoKitties. Again, this algorithm is written in Solidity and a smart contract. Smart contracts should output the same result if the input is the same. So, smart contracts cannot use a random number generator.  Therefore, we have to consider the system without a random number.

\begin{figure}
\centering
\includegraphics[width=8cm]{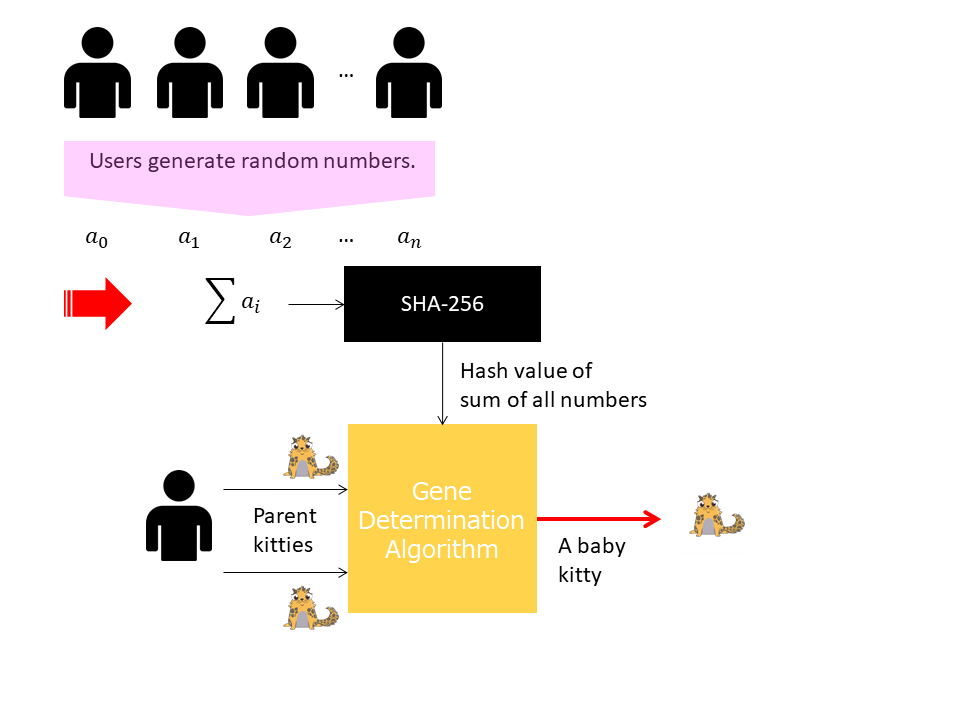}
\caption{Countermeasure's Overview} \label{fig5}
\end{figure}

To add an external randomness source in deciding the gene, we suggest that the users jointly create a random number as an input to a hash function (e.g., SHA-256) instead of creating value from a block on the blockchain.  For instance, when someone does breeding, everyone chooses a random number. The sum of them which users submit will be the input to a secure hash function. 
As all users cannot predict the input, its output is also unpredictable. This game's usability will not be compromised if a tool automatically generates and sends a random number every time requested. 
It provides fairness to the market since every user can have a chance to get a high-value ERC-721 token and trade it to Ethereum. We show this system's overview in Fig \ref{fig5}.

\subsection{The readability of Solidity source code}
We explained that if a player does not understand the gene determination algorithm, he may not get a valuable kitty from kitties he has in Section 3.2. For example, though he has a Diamond kitty, he could not inherit its Cattribute to its kitties. He will not make Ethereum because the kitty without Family Jewels is expected not popular. On the other hand, if another player understands the algorithm, he will create expensive kitties. We expect that there are significant gaps to opportunities to gain profit between knowledgable players and not. In order to make CryptoKitties fair, this game should fix so that all users can understand its gene algorithm. However, we insist that CryptoKitties does not achieve it. The readability of a programming language is not high, and it is estimated that more people cannot read it than those who can.

We assert that CryptoKitties needs a system so that more people can figure it out. These game managers should prepare not only to source code but also a diagram, flowchart, language description, and so on. It is expected that a variety of explanatory methods will reduce the information asymmetry about the gene determination algorithm, and this game will be satisfied with Condition No.2.

\subsection{Auction in CryptoKitties}
 
We show that when a seller and a bidder are colluding, it is difficult for other players to get the seller's kitty in Section 3.3. Though other players will not create this kitty's children, the bidder can. The seller gains profit by selling the kitty, and the bidder also earns Ethereum by selling child kitties of his kitty. But, other players have no opportunities to gain profit from the kitty. We argue that CryptoKitties does not provide an environment that every user has a chance to gain profit equally. This problem stems from the fact that the time between the auction starts and the winning bid is too short.

CryptoKitties should create an auction where everyone has an opportunity to trade. One countermeasure is that supposing it allows a certain amount of time, one hour, for example, between when a player puts up an item and when it becomes available for bidding. During this period, many people will be able to grasp the kitty the seller exhibits and think whether each of them should bid or not. We hope that it enhances to give all users a chance to get the kitty and make money.

%% file: 5_future_work.tex
\section{Future Work}
\subsection{Proof of Stake}

In the Ethereum environment, there is an assumption that no user has a large enough Ethereum to impact the market significantly. The symbol of this is Proof-of-Stake (PoS). Suppose hypothetically; some users had a substantial amount of Ethereum that could change the market value significantly. This situation is not satisfied with Condition No.4. In that case, it is quite likely that those users would be able to get the right to create blocks many times. There is no bias towards users' chance in terms of a decentralized system, which means to have the right to create. However, there is no countermeasure to prevent this from happening, and playability could be lost considerably.

\subsection{Secret trading}
\begin{figure}
\centering
\includegraphics[width=10cm]{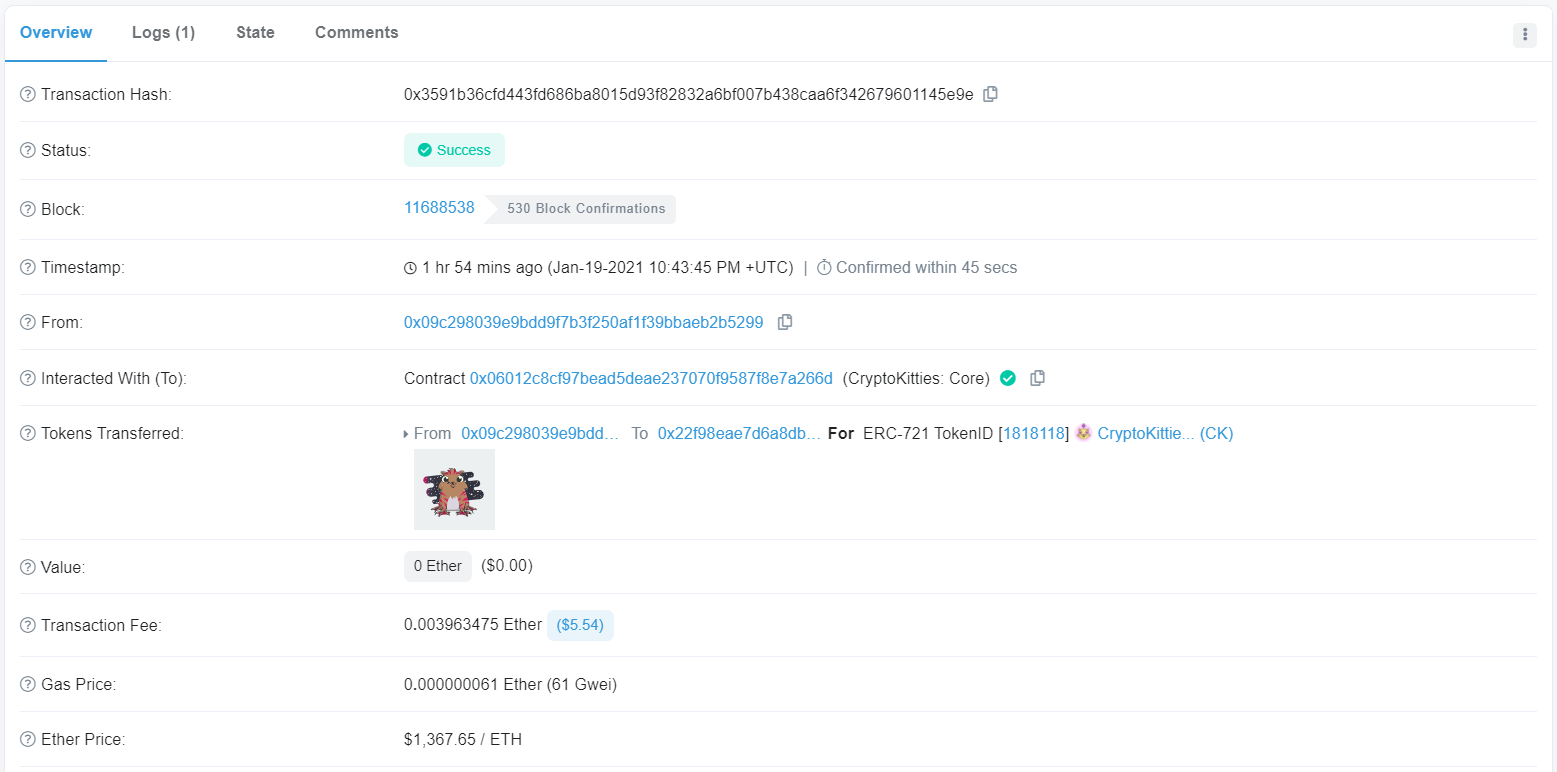}
\caption{One example of suspicious transactions} \label{fig3}
\end{figure}

We state that transactions can only be conducted through the auction market. However, is this really the case? We find some suspicious transactions, as shown in Fig.\ref{fig3} \cite{suspicious}. They traded kitties for nothing and without going through an auction possibly. We can not figure out how they traded. If they really sold using other methods, this game does not meet condition 3 and will be reversed. This game should also take measures for such cases.

\subsection{Blockchain anonymity}
Ethereum's blockchain is guaranteed to be anonymous. The blockchain address cannot be tied to who the actual user. However, there is one problem that arises from this. That is insider trading. Insider trading is that a person who has inside information about a company buys or sells shares before the information about the material fact is made public. Or, this company's employees trade them. In order to discover these transactions, it is necessary to know who made them. Unfortunately, it is so difficult to find them in the blockchain environment because of blockchain anonymity. If its developer takes part in this game and earns a large Ethereum, we will not see his illegal activity. Thus, CryptoKitties should prove that they are not doing this.

\subsection{Fee of transaction and breeding}
We argue that Gas and breeding fee are an obstacle to the motivation to trade for the average user. Transaction fee, GAS, is a miner's motivation because if he has a right to create a block, all GAS in the block will be his income. Miners tend to select transactions whose GAS are high. Therefore, wealthy users' transactions are apt to be approved since they pay high GAS. Besides, the Breed fee, currently 0.008 ETH, is CryptoKitties managers' income. When a player tries to breed his kitties, he pays Ethereum in exchange for a new kitty. Users who can afford Ethereum can breed or exchange kitty without hesitation. In contrast, for other players, their action count is limited. So, they have fewer opportunities to get good kitties and gain profit than rich people. In other words, these fees do not help conditions No.1 and No.5. One countermeasure we think is decreasing the breeding fee so that more players give birth to baby kitties many times.

%% file: 6_conclusion.tex
\section{Conclusion}

We found that CryptoKitties is not satisfied with four conditions that fair markets should keep. The gene determination algorithm which affects the ERC-721 token value is not satisfied with two conditions. We show that we can predict we will produce a token. In the environment we set, a player can not gain profit when he creates and sells a cheap token because of the breeding fee. So, he needs to sell a token whose value is more than the breeding fee. Since we can know which token we will make, we know tokens that can produce high-value ones. Thus, rich players are more easily to get lucrative limited tokens. Besides, people who can understand this algorithm are limited because the only Solidity expresses this algorithm. Users who know this bias have a significant advantage in playing CryptoKitties. Also, there is a problem with its trading market. When two players are colluding, it will be possible to attack them by not allowing others to trade with them.

We also mention the countermeasures of these problems. In the case of the gene algorithm, it needs an unpredictable system. Since this problem stems from seeing the input of the hash value it is using; we propose that all users determine its input. Moreover, describing these mechanisms in other than a programming language will reduce the information asymmetry. Finally, as one cause is that from the auction starts to the winning bid is too short, one method would be noticing what kitties have been put up for auction and allow a certain amount of time before the bidding starts. We propose these improvements, but there are other problems in CryptoKitties.

